\documentclass{jpsj3}

\usepackage{color}

\def\gsim{\mathop {\vtop {\ialign {##\crcr 
$\hfil \displaystyle {>}\hfil $\crcr \noalign {\kern1pt \nointerlineskip } 
$\,\sim$ \crcr \noalign {\kern1pt}}}}\limits}
\def\lsim{\mathop {\vtop {\ialign {##\crcr 
$\hfil \displaystyle {<}\hfil $\crcr \noalign {\kern1pt \nointerlineskip } 
$\,\,\sim$ \crcr \noalign {\kern1pt}}}}\limits}

\title{
On Pulsed Magnetic Field Measurements of Enhanced Sommerfeld Constant $\gamma$ and $A$-Coefficient around 
Metamagnetic Transition in UTe$_2$
}

\author{Kazumasa Miyake}

\inst
{
Center for Advanced High Magnetic Field Science, Osaka University, Toyonaka, 
Osaka 560-0043, Japan 
}

\recdate
{
\today
}

\abst
{
It is discussed from theoretical viewpoint 
why the aspects of enhancements in the $A$-coefficient in the resistivity\cite{Knafo} and  
the Sommerfeld constant $\gamma$\cite{AtsushiMiyake,Imajo} observed in UTe$_2$ 
around the metamagnetic transition at $H=H_{\rm m}$ are so different.  
In particular, the discussions are focused on the reason why the $H$ dependence of the $A$-coefficient reported 
in Ref. \citen{Knafo} and $\gamma$ reported in Ref. \citen{AtsushiMiyake} is almost symmetric 
around $H=H_{\rm m}$, while $\gamma(H)$ at $H\gsim H_{\rm m}$, reported in Ref. \citen{Imajo}, 
is considerably suppressed compared to that at $H\lsim H_{\rm m}$. 
A key point to solve this seeming paradox is to take properly 
into account the difference in time scales of probes used in those measurements of 
pulsed magnetic field method with different time scales. Another crucial effect is the 
time dependent {\it non-uniform} magnetic fluctuations remaining in a certain period of time 
after the {\it first-order} metamagnetic transition has occurred.   
Indeed, the time scales of former measurements in Refs. \citen{Knafo,AtsushiMiyake} are far shorter 
than that of the latter one in Ref. \citen{Imajo}, which can solve the above paradox.

}

\begin{document}
\sloppy
\maketitle
\section{Introduction} 
The Sommerfeld constant $\gamma$ and the $A$-coefficient in the temperature dependence of 
the resistivity  in the low temperature limit are crucial physical quantities in strongly correlated 
Fermi liquid metals. It is well recognized that both $\gamma$ and $A$ are highly enhanced 
in strongly correlated metals but the ratio $A/\gamma^{2}$ remains as the quasi-universal one  
from experimental  results~\cite{KW} and theoretical arguments~\cite{MV}.  
It was a mystery in early stage of research in UTe$_2$ that both quantities exhibit sharp 
increase around the {\it first}-order metamagnetic transition under the magnetic field along $b$-axis. 
One of possible answer was proposed in Ref. \citen{KazumasaMiyake} to explain such increase just at 
the metamagnetic field $H_b=H_{\rm M}$ and the fact that the ratio $A/\gamma^{2}$ is given approximately 
by that of Kadowaki-Woods~\cite{KW}. 
On the other hand, the $H_{b}$ dependence of  around $H_{b}=H_{\rm M}$ exhibits rather different 
aspects depending on details of experimental probes. The purpose of the present note is to clarify 
the origin of this fact by examining the character of those probes.

\section{Summary of Experimental Facts}
In this section, we summarize the characteristics of experimental facts of the metamagnetic 
transition observed in UTe$_2$.
\subsection{Variations of the splitting width of the matamagnetic fields of up-sweep and down-sweep processes}
In this subsection, we discus the aspect of the splitting of the metamagnetic fields, $H_{\rm m}$'s, for 
up-sweep and down-sweep processes.  
First, let us denote the splitting of the metamagneitc fields for up-sweep ($H_{\rm m}^{\rm up}$) and 
down-sweep ($H_{\rm m}^{\rm down}$) as $\Delta H_{\rm m}\equiv H_{\rm m}^{\rm up}-H_{\rm m}^{\rm down}>0$. 
The results for these splittings reported in Refs. \citen{Knafo} and \citen{AtsushiMiyake} are  
\begin{eqnarray}
\Delta H_{\rm m}^{\rm K}\simeq 0.53 {\rm T},
\label{Knafo}
\end{eqnarray}
and 
\begin{eqnarray}
\Delta H_{\rm m}^{\rm M}\simeq 0.36 {\rm T}, 
\label{AtsushiMiyake}
\end{eqnarray}
respectively. 
On the other hand, that reported in Ref. \citen{Imajo} is 
\begin{eqnarray}
\Delta H_{\rm m}^{\rm I}\simeq 0.10 {\rm T}.
\label{Imajo}
\end{eqnarray}

\subsection{Variations of the rate of change of the magnetic field at the metamagnetic transition in 
up-sweep process}
Corresponding to the variations of $(\Delta H_{\rm m})$'s discussed in the previous subsection, 
the rate of change of the magnetic field, $(dH/dt)|_{H=H_{\rm m}^{\rm up}}$'s, at the metamgnetic field of the 
up-sweep process exhibits large variations as follows. 
The rates reported in Refs. \citen{Knafo,Knafo2} and \citen{AtsushiMiyake,AtsushiMiyake2}  are 
\begin{eqnarray}
\left(\frac{dH}{dt}\right)_{H=H_{\rm m}^{\rm up}}^{\rm K}\gsim \frac{68-34.9}{(50/2)}\,\dot=\,1.3\,{\rm T/msec},
\label{Knafo5}
\end{eqnarray}
and 
\begin{eqnarray}
\left(\frac{dH}{dt}\right)_{H=H_{\rm m}^{\rm up}}^{\rm M}\simeq \frac{70-30}{(20/2)}\,\dot=\,4.0\,{\rm T/msec},
\label{AtsushiMiyake5}
\end{eqnarray}
respectively. 
On the other hand, that in Ref. \citen{Imajo} is far smaller than those given by Eqs. (\ref{Knafo5}) and 
(\ref{AtsushiMiyake5}) as 
\begin{eqnarray}
\left(\frac{dH}{dt}\right)_{H=H_{\rm m}^{\rm up}}^{\rm I}\simeq \frac{40-36.1}{(14\times10/2)}\,\dot=\,5.6\times 10^{-2}\,{\rm T/msec}.
\label{Imajo5}
\end{eqnarray}

%

\subsection{Variations of the time interval between the matamagnetic transitions of up-sweep and 
down-sweep processes}
Complementarily to the variations of $(\Delta H_{\rm m})$'s discussed in Subsec. 1.1, 
the time intervals $(\Delta t)$'s between two metamagnetic transitions, i.e., those corresponding to down-sweep 
and up-sweep processes, show large variations as follows.   
The intervals $(\Delta t)$'s reported in Refs. \citen{Knafo,Knafo2} and \citen{AtsushiMiyake,AtsushiMiyake2}  are 
\begin{eqnarray}
(\Delta t)^{\rm K}\simeq 50\,{\rm msec},
\label{Knafo4}
\end{eqnarray}
and 
\begin{eqnarray}
(\Delta t)^{\rm M}\simeq 20\,{\rm msec},
\label{AtsushiMiyake4}
\end{eqnarray}
respectively. 
On the other hand, that reported  in Ref. \citen{Imajo} is far longer than those given by 
 Eqs. (\ref{Knafo4}) and (\ref{AtsushiMiyake4}) as 
\begin{eqnarray}
(\Delta t)^{\rm I}\simeq 14\times 10\,{\rm msec}.
\label{Imajo4}
\end{eqnarray}
Note, furthermore, that the measurements of the Sommerfeld constant $\gamma$ in Ref. \citen{Imajo2} 
were performed by the quasi-adiabatic method, in which each measurement at specific strength of magnetic field 
has been performed under the {\it almost} static magnetic field during about 400 msec 
which is longer  moreover than $(\Delta t)^{\rm I}\simeq 140\ {\rm msec}$ [Eq. (\ref{Imajo4})].

\section{Interpretation of Experimental Facts}
In this section, we discuss the meaning of experimental facts shown in the previous section.  
The facts shown in Subsects. 1.1 and 1.2 imply that the metamagnetic transition is controlled 
not only by the balance of the static free energy but it occurs through the metastable states 
whose life time must depend on the rates of magnetic field change. 
In general, an ideal {\it first-order} metamagnetic transition under the adiabatic process occurs  
when the free energies of the two stable states with $M<M_{\rm m}$ and the unstable state $M>M_{\rm m}$ 
coincide with each other as the magnetic field $H$ increases adiabatically through $H=H_{\rm m}$. 
However, in reality, the state at $H<H_{\rm m}$ continues to stay as the metastable one even at $H>H_{\rm m}$ 
{\it in a certain period of time}, depending on the rate of increase of $H$ unless the instability point is reached 
where the {\it sponodal} like process is forcibly started to occur, 
instantaneously giving  rise to the discontinuous transition.  
Indeed, the correspondence between variations of $\Delta H_{\rm m}$'s shown in Eqs. (\ref{Knafo}), 
(\ref{AtsushiMiyake}), and (\ref{Imajo}), 
and those of  $(dH/dt)|_{H=H_{\rm m}^{\rm up}}$'s given approximately by Eqs. (\ref{Knafo5}), 
(\ref{AtsushiMiyake5}), and (\ref{Imajo5}), respectively, would support this point of view.  
Namely, this circumstantial evidence suggests the importance to take into account 
the time dependent process beyond the competition among static free energies, such as the 
extended Landau-type free energy used for understanding 
the enhancements of  the Sommerfeld constant $\gamma$ and the $A$-coefficient {\it just at} the 
metamagnetic transition discussed in Ref.\ \citen{KazumasaMiyake}.  

\subsection{Free energy shift due to thermodynamic magnetic fluctuations}
In this subsection, we discuss a possible origin of excess free energy gain due to magnetic fluctuations 
beyond the static free energy of the extended Landau-type free energy, used in Ref. \citen{KazumasaMiyake}. 
 
According to a general principle of statistical physics,~\cite{LandauLifshitz} 
the free energy is influenced also by the thermodynamic fluctuations, i.e., {\it non-uniform} magnetic fluctuations 
in the present case. 
Namely, the shift of the free energy $\Delta F$ due to the fluctuations, under fixed temperature and volume, 
is given as follows:
\begin{eqnarray}
\Delta F=-k_{\rm B}\log\Delta Z_{\rm fl},
\label{FEFL1}
\end{eqnarray}
where $k_{\rm B}$ is the Boltzmann constant and the partition function $\Delta Z_{\rm fl}$ is defined as 
\begin{eqnarray}
\Delta Z_{\rm fl}=\Pi_{\bf q}\int d{\boldsymbol M}_{\bf q}^{\prime}d{\boldsymbol M}_{\bf q}^{\prime\prime}
{\rm exp}\left[-(\eta+Aq^{2})|{\boldsymbol M}_{\bf q}|^{2}/k_{\rm B}TN_{\rm F}^{*}\right],
\label{FEFL2}
\end{eqnarray}
where ${\boldsymbol M}_{\bf q}\equiv {\boldsymbol M}_{\bf q}^{\prime}+i{\boldsymbol M}_{\bf q}^{\prime\prime}$.  
Substituting $\Delta Z_{\rm fl}$ [Eq. (\ref{FEFL2})] into Eq. (\ref{FEFL1}), $\Delta F$ is expressed as 
\begin{eqnarray}
& &
\Delta F=-k_{\rm B}\sum_{\boldsymbol q}\log
\left\{
\int d{\boldsymbol M}_{\bf q}^{\prime}d{\boldsymbol M}_{\bf q}^{\prime\prime}
{\rm exp}
\left[-(\eta+Aq^{2})|{\boldsymbol M}_{\bf q}|^{2}/k_{\rm B}TN_{\rm F}^{*}
\right]
\right\}
\nonumber
\\
& &
\qquad
=-k_{\rm B}\sum_{\boldsymbol q}\log\left(\frac{\pi k_{\rm B}TN^{*}_{\rm F}}{\eta+Aq^{2}}\right),
\label{FEFL3}
\end{eqnarray}
which is negative in the thermodynamic limit, i.e., $N\to \infty$. 
Note that the form of Eq.\ (\ref{FEFL3}) reflects the dynamical spin susceptibility of ferromagnetic 
fluctuations discussed in Ref.\ \citen{Moriya}, i.e., 
\begin{eqnarray}
\chi_{\rm s}({\boldsymbol q},{\rm i}\omega_{m})
=\frac{qN_{\rm F}^{*}/C}
{\omega_{\rm s}(q)+|\omega_{m}|}, \quad{\hbox{for
$q<q_{\rm c}\sim p_{\rm F}$}}, 
\label{SF1}
\end{eqnarray}
where $N_{\rm F}^{*}$ is the density of states (DOS) of the quasiparticles {\it per spin} at the Fermi level 
and $\omega_{\rm s}(q)$ is defined as 
\begin{eqnarray}
\omega_{\rm s}(q)\equiv \frac{q}{C}(\eta+Aq^{2}),
\label{SF2}
\end{eqnarray}
where $\eta$ parameterizes the closeness to the ferromagnetic criticality. 
Note that UTe$_2$ is considered to be located near the ferromagnetic criticality as discussed in 
Refs. \citen{Ran,Aoki1,Sundar}.  Therefore, the contribution of  $\Delta F$ is considered to have a certain effect 
on the metamagnetic transition as discussed below.     

\subsection{Condition for metamagnetic transition in pulsed magnetic field experiments}
In this subsection, we discuss the aspect of the metamagnetic transition in the up-sweep process 
in the pulsed magnetic field experiments. 

\begin{figure}[h]
\begin{center}
\rotatebox{0}{\includegraphics[width=0.7\linewidth]{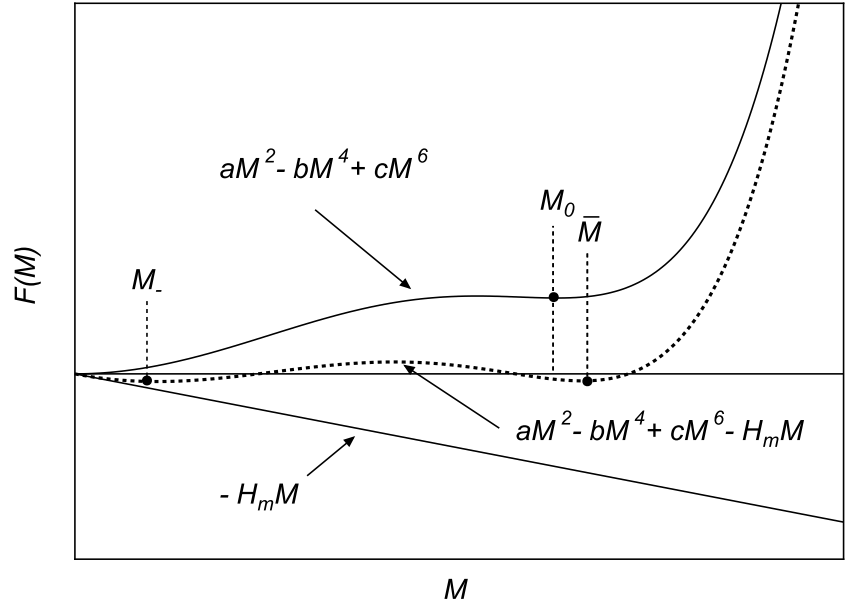}}
\caption
{Schematic behavior of free energies, appearing in Eq.\ (\ref{FE1}), 
as a function of the uniform magnetization. Solid curve and solid line 
represent the first three parts, free energy $F_{0}(M)$ [Eq.\ (\ref{FE1})] without the magnetic field $H$ and the Zeeman 
energy at the metamagnetic field $H_{\rm m}$, respectively. The dotted curve represents the free energy 
$F_{0}(M)-H_{\rm m}M$ which has two degenerate minima at $M=M_{-}$ and ${\bar M}$.    
}
\label{Fig:FreeEnergy}
\end{center}
\end{figure}

In Ref. \citen{KazumasaMiyake}, the extended Landau-type free energy $F(M;H)$ under the magnetic field $H$ 
was adopted as  
\begin{eqnarray}
F(M;H)\simeq aM^{2}-bM^{4}+cM^{6}-HM, 
\label{FE1}
\end{eqnarray}
where coefficients $a$, $b$ and $c$ are assumed to be positive, and tuned to recover the 
gross characters of the metamagnetic transition and concomitant anomalies of the Sommerfeld constant $\gamma$ 
and the $A$-coefficient {\it just} at $H=H_{\rm m}$. 
The condition for the metamagnetic transition discussed in Ref. \citen{KazumasaMiyake} was given by 
the condition that the free energies at $M=M_{-}$ and $M={\bar M}$ coincide, i.e., 
$F(M_{-})=F({\bar M})$, as schematically shown in Fig.\ \ref{Fig:FreeEnergy}. 
However, as noted above, the excess contribution $\Delta F$ [Eq.\ (\ref{FEFL3})]  
beyond $F(M;H)$ [Eq.\ (\ref{FE1})] is crucial to determine the {\it ideal} or {\it adiabatic} metamagnetic transition. 
Namely, its condition in the up-sweep process of the pulsed magnetic field experiment should have been replaced by  
\begin{eqnarray}
F(M_{-};H_{\rm m}^{\rm up})+\Delta F(H_{\rm m}^{\rm up})=F({\bar M};H_{\rm m}^{\rm up}),
\label{Transition1}
\end{eqnarray}
where $\Delta F(H_{\rm m}^{\rm up})$ expresses the contribution from the {\it non-uniform} magnetic fluctuations 
around $M=M_{-}$, one of the degenerate local minima, while such a contribution from the fluctuations around 
$M={\bar M}$ has been discarded because it stays as a virtual one before the {\it adiabatic} 
metamagnetic transition occurs, i.e., $H\le H_{\rm m}^{\rm up}$.

Furthermore, the stable state at $H\le H_{\rm m}^{\rm up}$ remains to be metastable before the instability 
point is reached and the {\it spinodal} like process is caused forcibly to reach the true stable higher field phase.  
However, according to the experimental facts shown in Subsects 1.1 and 1.2, the observed metamagnetic 
fields $H_{\rm m}^{\rm up}$'s are distributed depending on the rates of increasing the magnetic field, implying that the 
{\it spinodal} point is not reached in actual experiments. This is because the {\it spinodal} point must be determined, 
independently of the rate of increasing the magnetic field $H$, by the condition that the following two relations 
hold simultaneously {\it for the first time} as increasing $M$ together with $H$: 
\begin{eqnarray} 
\frac{\partial}{\partial M}\bigl[F(M;H_{\rm sp}^{\rm up})+\Delta F(H_{\rm sp}^{\rm up})\bigr]=0,
\label{spinodal0}
\end{eqnarray}
and 
\begin{eqnarray} 
\frac{\partial^{2}}{\partial^{2}M}\bigl[F(M;H_{\rm sp}^{\rm up})+\Delta F(H_{\rm sp}^{\rm up})\bigr]=0.
\label{spinodal1}
\end{eqnarray}
Here, $H_{\rm sp}^{\rm up}$ is the magnetic field of starting the {\it spinodal} like transition in the up-sweep 
process. Note that  $H_{\rm m}^{\rm up}$ must approach $H_{\rm sp}^{\rm up}$ from the lower side in the 
limit of $(dH/dt)_{H=H_{\rm m}^{\rm up}}\rm =\infty$. 
Although the precise mechanism of the {\it real}  transition from the metastable state to the stable one at 
$H\ge H_{\rm m}^{\rm up}$ 
has not been clarified yet, as a working hypothesis, we assume the existence of some mechanism 
to cause this transition  on the basis of experimental observations above.     

Needless to say, the manner of the {\it inverse} metamagnetic transition in the down-sweep process is similar to 
that in the up-sweep but in inverse way. Namely, the conditions [Eqs. (\ref{Transition1}), (\ref{spinodal0}), 
and (\ref{spinodal1})] 
are replaced by 
\begin{eqnarray}
F({\bar M};H_{\rm m}^{\rm down})+\Delta F(H_{\rm m}^{\rm down})=F(M_{-};H_{\rm m}^{\rm down}),
\label{Transition2}
\end{eqnarray}
\begin{eqnarray} 
\frac{\partial}{\partial M}\bigl[F(M;H_{\rm sp}^{\rm down})+\Delta F(H_{\rm sp}^{\rm down})\bigr]=0,
\label{spinodal0down}
\end{eqnarray}
and 
\begin{eqnarray} 
\frac{\partial^{2}}{\partial^{2}M}\bigl[F(M;H_{\rm sp}^{\rm down})+\Delta F(H_{\rm sp}^{\rm down})\bigr]=0,
\label{spinodal2down}
\end{eqnarray}
respectively. 

\subsection{Origin of difference of anomalies in $\gamma$ and $A$ coefficients around the metamagnetic transition 
in pulsed magnetic field measurements} 
Finally, in this subsection, we propose a possible scenario to understand why the difference arises in   
anomalies of $\gamma$ and $A$-coefficient around the {\it first-order} metamagnetic transition 
in pulsed magnetic field measurements.  Namely, the discussions are focused on the reason why the $H$ dependence 
of the $A$-coefficient\cite{Knafo} and the Sommerfeld constant $\gamma$\cite{AtsushiMiyake} is 
almost symmetric around $H=H_{\rm m}$, 
while $\gamma(H)$ at $H\gsim H_{\rm m}$, reported in Ref. \citen{Imajo}, is considerably suppressed 
compared to that at $H\lsim H_{\rm m}$. 

A hint to solve this paradox must be the fact that the metamagnetic transition is determined not only by 
the extended Landau-type free energy [Eq. (\ref{FE1})] for the {\it uniform} ferromagnetic 
order parameter $M$, but is influenced by the excess free energy $\Delta F$ [Eq. (\ref{FEFL3})] 
arising from the {\it non-uniform} magnetic fluctuations.  Indeed, this aspect was pointed out in the previous subsection 
in relation to the condition [Eq. (\ref{Transition1})] for the {\it ideal} and {\it adiabatic} metamagnetic transition.  

Another circumstantial aspect of the metamagentic transition under the pulsed magnetic field measurements is that 
the actual metamagnetic field $H_{\rm m}^{\rm up}$'s in the up-sweep process depend on the rate of increasing 
magnetic fields as suggested by experimental facts shown in Subsects. 1.1 and 1.2. 
This aspect implies that the metamagnetic transition in question is quite a time dependent phenomenon. 

The most crucial fact would be that the above-mentioned {\it non-uniform} magnetic fluctuations remain 
as non-vanishing in a certain period of time $(\Delta t)_{\rm fl}$ 
even after the metamagnetic transition, i.e., in the region $H\gsim H_{\rm m}^{\rm up}$.  
Note that these magnetic fluctuations are described by the dynamical susceptibility [Eq. (\ref{SF1}) with Eq. (\ref{SF2})] 
which are the {\it same} fluctuations giving rise to the enhancements of $\gamma$ and $A$-coefficient in the region 
$H\le H_{\rm m}^{\rm up}$. Therefore, such enhancements of $\gamma$ and $A$-coefficient should remain 
even in the region $H\gsim H_{\rm m}^{\rm up}$ in the period $(\Delta t)_{\rm fl}$ in general. 
Then, the question is how long $(\Delta t)_{\rm fl}$ is.  
Although the reliable theoretical estimation of $(\Delta t)_{\rm fl}$ has not been available, its possible range 
can be inferred so as to explain consistently experimental observations shown in Sect. 1.  


The time intervals $(\Delta t)$'s between the metamagnetic transitions of up-sweep and down-sweep processes 
in Refs. \citen{Knafo} and \citen{AtsushiMiyake} are given by Eqs. (\ref{Knafo5}) and (\ref{AtsushiMiyake5}), 
respectively. The almost symmetric behaviors of $\gamma$ and $A$-coefficient imply that 
$(\Delta t)_{\rm fl}$ would be comparable to or longer than 
$(\Delta t)^{\rm K}\simeq 50 {\rm msec}>[(\Delta t)^{\rm M}\simeq 20 {\rm msec}]$.  
Note that $\gamma$ reported in Ref. \citen{AtsushiMiyake} was estimated by using the Maxwell relation 
between M and the entropy S, i.e., $(\partial S/\partial H)_{T}=(\partial M/\partial T)_{H}$, so that 
the duration time of measurement is about $(\Delta t)^{\rm M}\simeq 20 {\rm msec}$ [Eq. (\ref{AtsushiMiyake5})],  
during of which the effect of the {\it non-uniform} magnetic fluctuations almost remains 
giving rise to a considerable enhancements of 
$\gamma$ and $A$-coefficient as really observed in Refs. \citen{Knafo} and \citen{AtsushiMiyake}. 
On the other hand, the specific heat measurements in Ref. \citen{Imajo} have been performed by the 
quasi-adiabatic method\cite{Imajo2}, whose duration time of {\it each} measurement at specific strength of 
magnetic field is about 400 msec which is far longer than $(\Delta t)_{\rm fl}$ estimated above.  
Therefore, the contribution from the remnant {\it non-uniform} magnetic fluctuations would be considerably suppressed 
in the high field region $H>H_{\rm m}^{\rm up}$ as observed in Ref. \citen{Imajo}. 

\subsection{Conclusion of Sect. 3} 
As discussed i the previous subsection, the seemingly conflicting aspects of the enhancement of 
the Sommerfeld constant $\gamma$ and the $A$-coefficient around the metamagnetic transition, 
which is summarized in the first paragraph of this section (Sect. 3), 
can be understood naturally by taking carefully into account the deference in the 
time dependence of  the pulsed magnetic field probes.

\section{Conclusion and Perspective}
A possible scenario has been proposed why the aspects of the enhancements of the Sommerfeld constant 
$\gamma$ anf the $A$-coefficient of the resistivity, near the {\it first order} metamagnetic transition of 
UT$_2$ observed by a series of pulsed magnetic field measurements, are rather different. 
It was crucial that the metamagnetic transition under the pulsed magnetic field measurements 
is influenced by the rate of changing the magnetic field.  
Therefore, it became indispensable to consider the effect of the {\it non uniform} and {\it time dependent}
magnetic fluctuations around the stationary point of the {\it static} Landau-type free energy for the {\it uniform} 
magnetization. 

In order to verify the present scenario, a series of  measurements of the $A$-coefficient are expected, 
with changing $(\Delta t)$, time interval between the metamagnetic transitions of up-sweep and down-sweep 
processes, to approach that used in Ref. \citen{Imajo}.

\section*{Acknowledgments}
The present note is written to answer the deep question posed by Jacques Flouquet whose query is gratefully 
acknowledged.  
The author is grateful to Atsushi Miyake and Shusaku Imajo for informative discussions on experimental 
aspects of the present problem.  A communication with William Knafo was also useful for reinforcing the 
idea of the present note.  Conversation with Masayuki Hagiwara about pulsed magnetic field measurement 
in general is acknowledged as useful to construct the scenario of the present note. 
This work is supported by JSPS-AKENHI (No. 17K05555).




\end{document}